\documentclass[11pt,a4paper]{article}
\pdfoutput=1

\usepackage{jheppub,bm,mathtools,xfrac}

\title{\boldmath Transverse-momentum-dependent  quark splitting functions  in $k_T$-factorization: real contributions  
}

\author[a]{Oleksandr~Gituliar,}
\author[b,c]{Martin Hentschinski,}
\author[a]{Krzysztof Kutak}
\affiliation[a]{Instytut Fizyki J\c{a}drowej Polskiej Akademii Nauk, Radzikowskiego 152, 31-342 Krak\'ow, Poland}
\affiliation[b]{Instituto de Ciencias Nucleares, 
Universidad Nacional Aut\'onoma de M\'exico,
Apartado Postal 70-543
M\'exico D.F. 04510 MX}
\affiliation[c]{Instituto de F\'isica y Matem\'aticas, Universidad Michoacana de San Nicol\'as de Hidalgo, Apartado Postal 2-82, Morelia, Michoacan 58040, MX}
\emailAdd{oleksandr.gituliar@ifj.edu.pl,\\hentschinski@correo.nucleares.unam.mx,\\krzysztof.kutak@ifj.edu.pl}

\toccontinuoustrue

\usepackage{slashed,subcaption,xspace}

\newcommand{\D}{\mathrm{d}}

\newcommand{\eps}{\epsilon}
\renewcommand{\sp}[2]{\,#1\!\cdot\!#2\,}

\newcommand{\Cf}{C_\mathnormal{F}}
\newcommand{\eqb}{\begin{equation}}
\newcommand{\eqe}{\end{equation}}
\newcommand{\be}{\begin{equation}}
\newcommand{\ee}{\end{equation}}

\newcommand{\kt}{\mathbf{k}}
\newcommand{\Pbb}{\mathbb{P}}
\newcommand{\qt}{\mathbf{q}}
\newcommand{\pt}{\mathbf{\tilde{p}}}
\newcommand{\qtt}{\mathbf{\tilde{q}}}
\newcommand{\mlb}{\begin{multline}}
\newcommand{\mle}{\end{multline}}
\newcommand{\Tr}{T_\mathnormal{R}}

\newcounter{askcounter}

\numberwithin{askcounter}{section}

\abstract{We calculate transverse momentum dependent quark splitting
  kernels $P_{gq}$ and $P_{qq}$ within $k_T$-factorization, completing
  earlier results which concentrated on gluon splitting functions
  $P_{gg}$ and $P_{qg}$. The complete set of splitting kernels is an
  essential requirement for the formulation of a complete set of
  evolution equations for transverse momentum dependent parton
  distribution functions and the development of corresponding parton
  shower algorithms.}

\begin{document}

\maketitle
\flushbottom

\section{Introduction}
The essential theoretical input for experimental findings at the Large
Hadron Collider are parton distribution functions (PDFs) which
describe momentum distributions of
partons in the colliding hadrons in the presence of a hard scale.  Together with factorization theorems
and hard coefficient functions, PDFs allow to predict new phenomena
and to describe existing data. A lot of recent activity in theory and
phenomenology of QCD is devoted to so called
transverse-momentum-dependent parton distribution functions (TMD PDFs)
and TMD factorization (for a review we refer the Reader to
\cite{Angeles-Martinez:2015sea}).  While a rigorous formulation of TMD
factorization, valid for all kinematic regions, is still to be
achieved (see e.g. \cite{Col11}), a definition of TMD parton
distributions is possible for specific regions of phase space, usually
characterized by a hierarchy of scales
\cite{Balitsky:2015qba,Dominguez:2011wm,Kotko:2015ura,Kovchegov:2015zha}.
One of those regions is the high-energy or small-$x$ limit of
perturbative QCD, characterized by the hierarchy
$\sqrt{s} \gg M \gg \Lambda_{\text{QCD}}$, where $\sqrt{s}$ denotes
the center-of-mass energy of the process, $M$ the hard scale of the
perturbative event, and $\Lambda_{\text{QCD}}$ the QCD characteristic
scale of the order of a few hundred MeV. The underlying theoretical
framework for TMD PDFs in this kinematic limit is usually referred to
as \mbox{$k_T$-factorization} or high-energy factorization \cite{Catani:1990eg}.  During the recent
years various hard processes, in particular those associated with the
forward region of LHC detectors, characterized by large rapidities,
have been studied within the \mbox{$k_T$-factorization} framework,
such as forward jet and forward $b$-jet production
\cite{Deak:2011ga,Deak:2010gk,Chachamis:2015ona} and forward
$Z$-production \cite{Dooling:2014kia,Hautmann:2012sh,vanHameren:2015uia}.
\\

In the following we are in particular interested in the evolution of
TMD PDFs, which depends on the parton's longitudinal momentum fraction
$x$, its transverse momentum $k_T$, and the external hard scale $M$.
An evolution equation which has these elements and is valid in angular
ordered phase space for gluon emission is provided by the
Ciafaloni-Catani-Fiorani-Marchesini (CCFM) equation
\cite{Cia87,Marchesini:1994wr,CFM90a,CFM90b}.  The key element of the
evolution kernel of the CCFM equation is the $P_{gg}$ splitting
function. At leading order it contains only the most singular pieces
at low $z \to 0$ and large $z \to 1$ and appropriate form factors
which resum virtual and unresolved real emissions in respectively low
and large $z$ regions.
\\

The CCFM equation is restricted to the resummation of purely gluonic
emissions.  In particular this implies that the large-$x$ behavior of
CCFM is not accurate and the formal large-$z$ limit of the CCFM
equation is incomplete, since it does not reduce to the matrix-valued
DGLAP evolution equations. One of the observations based on the
Monte-Carlo implementation \cite{Jung:2010si} of the CCFM equation is
that the lack of such contributions leads indeed to non-negligible
effects.  Performing a fit to the proton structure function $F_2$ at
both large and small $x$, it is likely that the gluon contribution is
enhanced in regions where quarks in the evolution would contribute.
While for inclusive observables, such as the structure function $F_2$,
the overall fit turns out to be satisfactory, see e.g.
\cite{Hautmann:2013tba}, the predictions based on the gluon density
are not satisfactory for exclusive observables, see e.g.
\cite{Deak:2010gk}. While it is difficult to pinpoint the exact reason
for this deficiency, DGLAP resummation definitely suggests that
decoupled evolution of quarks and gluons is insufficient. This is
further supported by application of the Kutak-Sapeta (KS) gluons
densities \cite{Kutak:2012rf,Kutak:2014wga} which account for quark
contribution in the evolution \cite{Kwiecinski:1997ee} and describes
production of dijets in p+p collisions at LHC reasonably well
\cite{Kutak:2012rf,vanHameren:2014ala}. In order to be able to apply
CCFM evolution successfully and to provide full parton shower
Monte-Carlo description within CCFM, the ultimate goal must be therefore to
arrive at a coupled system of equations which in turn requires a full
set of $k_T$-dependent splitting functions \cite{Jung:REF}.
\\

To arrive at a complete and consistent set of evolution equations, it
is further necessary to include --- apart from the quark splitting
functions $P_{gq}$ and $P_{qq}$ --- non-singular terms of the $P_{gg}$
splitting function since these corrections are of the same order
beyond leading order (LO) CCFM, i.e. beyond large- and small-$z$
enhanced contributions.  Note that in \cite{Jung:2010si} it has been
observed that inclusion of non-singular pieces of the DGLAP gluon
splitting function into CCFM evolution strongly affects the solution
of the evolution equation.  One may therefore conclude that the effect
of quarks in the evolution will be similarly significant.

A first step into this direction has undertaken in
\cite{Hautmann:2012sh}, where the TMD gluon-to-quark splitting kernel
$P_{qg}$ obtained in \cite{CH94} has been used to define a TMD
sea-quark density within $k_T$-factorization.  In the following we
extend this result by calculating as a start the unintegrated real
emissions kernels for quark-to-quark and quark-to-gluon splitting
functions.

From a technical point of view the determination of TMD splitting
kernels is based on a generalization of the high energy factorization
approach of Catani and Hautmann \cite{CH94}, which itself is based on
the formulation of DGLAP evolution in terms of a
two-particle-irreducible (2PI) expansion \cite{CFP80} (for overview and  recent applications of the method see \cite{Jadach:2011kc,Jadach:2011cr,Gituliar:2014mua,Gituliar:2014eba})). To guarantee
gauge invariance in presence of off-shell particles we follow the
proposal made in \cite{Hautmann:2012sh} and make use of the effective
action formulation of the high energy factorization in terms of
reggeized quarks and gluons \cite{Lipatov:2000se,Lipatov:1995pn}. In
the case of the gluon channel, consistency of this formalism has been
verified up to the 2-loop level through explicit calculations of the
higher-order corrections
\cite{Hentschinski:2011tz,Hentschinski:2011xg,Chachamis:2012gh,Chachamis:2012cc,Chachamis:2013hma}
and has been recently used to determine the complete next-to-leading
order corrections to the jet-gap-jet impact factor
\cite{Hentschinski:2014lma,Hentschinski:2014bra,Hentschinski:2014esa}.
\\

The outline of the present paper is the following: in
Sec.~\ref{sec:method} we give a comprehensive review of the results of
\cite{Hautmann:2012sh} and explain the strategy of our
calculations. In Sec.~\ref{sec:splitt-kern-from} we determine TMD
splitting functions working in the physical light-cone gauge,
following closely the setup of~\cite{CH94, CFP80}. In
Sec.~\ref{sec:unint-splitt-funct} we provide an extension of this
formalism which makes the gauge invariance of our result explicit,
despite of the presence of the off-shell legs in the matrix elements.
In Sec.~\ref{sec:conclusions-outlook} we summarize our results and
discuss directions for future
research.

\section{The method}\label{sec:method}

We start our presentation with a short review of the results of
\cite{CH94,Hautmann:2012sh} which allowed for the definition of the
TMD $P_{qg}$ splitting function and eventually of the sea-quark
density. The derivation follows two steps:
\begin{itemize}
\item[a)] in \cite{CH94} a TMD splitting function has been determined
  to construct a high-energy  resummed collinear sea-quark density. Its
  derivation is based on the two-particle-irreducible (2PI) expansion
  of \cite{CFP80}. To identify the TMD splitting function, one employs
  high-energy  factorization of the 2PI kernel into a TMD dependent
  gluon-to-quark splitting, i.e. the TMD splitting function, and
  the BFKL Green's function, which achieves a resummation of small $x$
  logarithms. To obtain the small-$x$ resummed sea-quark
  distribution, the TMD splitting function is combined with the BFKL
  Green's function and integrated over the transverse sea-quark
  momentum, following the conventions of \cite{CFP80}.
\item[b)] in \cite{Hautmann:2012sh} the limitation to the
  transverse-momentum-independent sea-quark distributions has been
  relaxed.  To ensure gauge invariance in the presence of off-shell
  splitting kernels, factorization of the process $qg^* \to qZ$ in the
  high-energy limit as realized by the reggeized quark formalism
  \cite{Bogdan:2006af,Lipatov:2000se} has been employed. Generalizing
  the reggeized quark formalism to finite energies, while taking care
  of maintaining gauge invariance, it was then possible to factorize
  the $qg^* \to qZ$ matrix element into a TMD coefficient $qq^* \to Z$
  and the TMD gluon-to-quark splitting function of
  \cite{CH94}. In particular, combining the TMD gluon-to-quark
  splitting function with the CCFM resummed TMD gluon distribution, a
  definition of a TMD sea-quark distribution has been achieved.
\end{itemize}
In the following we generalize these results to the quark-to-gluon and
quark-to-quark splittings, employing the two-step procedure outlined
above: we first define the splitting functions within the 2PI
expansion of \cite{CFP80, CH94} and then generalize our results to the fully off-shell splittings with full dependence on the transverse momentum. Before
turning to the derivation we would like to point out a
slight extension of the result of \cite{Hautmann:2012sh}. While
\cite{Hautmann:2012sh} concentrates on factorization of a particular
process, namely $qg^* \to qZ$, one can easily show that
the resulting matrix elements and TMD splitting functions are process-independent. To this end we recall the details of the high-energy 
factorization of the $qg^* \to qZ$ matrix element: 
\begin{figure}[th]
  \centering
  \parbox{4cm}{\center \includegraphics[width=4cm]{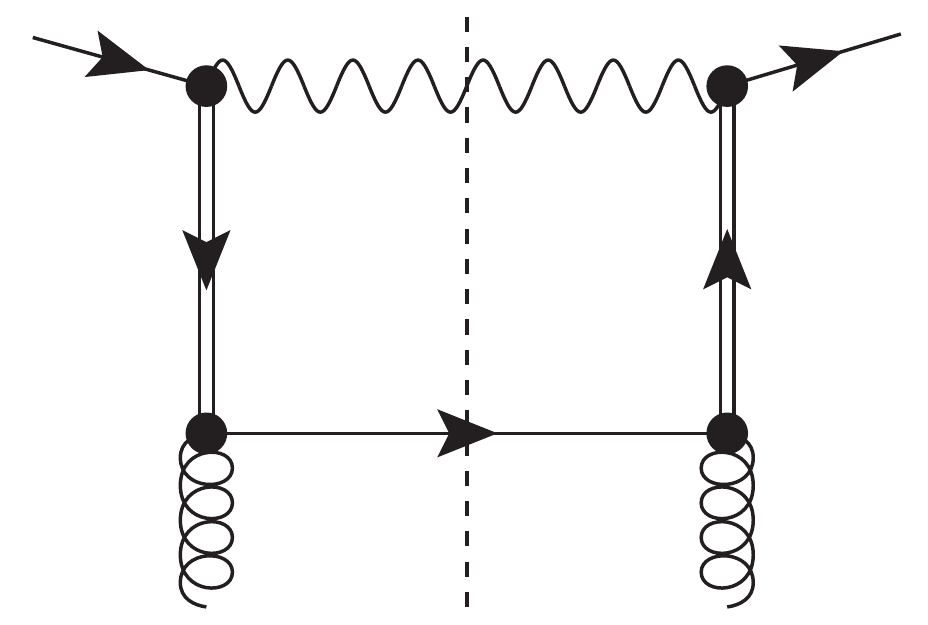}}
  \caption{\it The $g^*q \to Zq$ process within the reggeized quark
    formalism. Double lines with arrow indicate the effective
    reggeized quark exchange in the $t$-channel.}
  \label{fig:DY_reggeized}
\end{figure}
within the reggeized quark formalism, the entire process is described
using a single diagram, Fig.~\ref{fig:DY_reggeized}, with the
$qq^* \to Z$ and $g^*q^* \to q$ sub-amplitudes connected by reggeized
quark propagators,
\begin{align}
  \label{eq:RQprop}
  \parbox{1cm}{ \includegraphics[height=2cm]{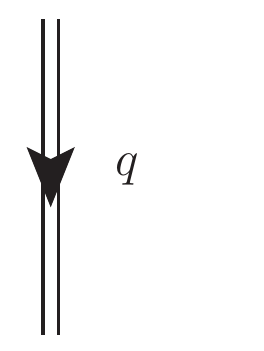}} &=  \frac{(\slashed{n} \slashed{p})_{\beta\alpha}}{2 p \cdot n} \cdot \frac{i \cdot \slashed{ q}}{ { q}^2+ i\epsilon }
&
\parbox{1cm}{ \includegraphics[height=2cm]{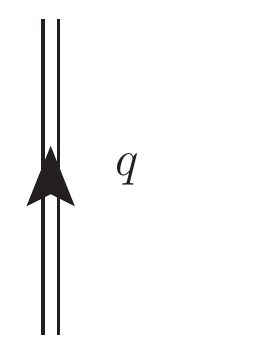}} &=  \frac{(\slashed{p} \slashed{n})_{\alpha\beta}}{2 p \cdot n} \cdot \frac{i \cdot \slashed{ q}}{ {q}^2+ i\epsilon }\,.
\end{align}
While in the strict high-energy  limit the $t$-channel four-momentum is
purely transverse, $q^2 = - {\bm q}^2$, generalizations to finite
energies require to keep the full momentum dependence. The momenta $p$
and $n$ are light-cone momenta $p^2 = n^2 = 0$ associated with the
almost light-like momenta of scattering particles normalized to
$2 p \cdot n = s$, with $s$ the center-of-mass energy of the hadronic
process. In e.g. deep-inelastic scattering, $n$ would be
associated with the virtual photon and $p$ with the probed
hadron. While (generalized) reggeized quark propagators carry at first
explicit spin indices and therefore correlate the $qq^* \to Z$ and
$g^*q^* \to q$ sub-amplitudes, it is possible to rewrite the high-energy  projectors for the cross-section using
\begin{align}
  \label{eq:1}
  \frac{ (\slashed{n}\slashed{p})_{\beta_1\alpha_1}  (\slashed{p}\slashed{n})_{\alpha_2\beta_2}}{p \cdot n} & = \slashed{p}_{\alpha_1\alpha_2}  \slashed{n}_{\beta_1\beta_2} +  (\gamma_5 \slashed{p})_{\alpha_1\alpha_2} (\gamma_5 \slashed{n})_{\beta_1\beta_2}.
\end{align}
For helicity independent input, the second term can be neglected and
one remains with the projector
$ \slashed{p}_{\alpha_1\alpha_2} \slashed{n}_{\beta_1\beta_2}$ which
then only contracts the Dirac indices of the $qq^* \to Z$ and
$g^*q^* \to q$ sub-amplitudes respectively and therefore leads to a
complete factorization of both processes.

\section{Splitting functions from the 2 PI expansion in the axial gauge}
\label{sec:splitt-kern-from}

The decomposition into 2PI diagrams as introduced in \cite{CFP80} is
based on the use of axial i.e. light-cone gauge, which allows
to analyze collinear singularities on the graph-by-graph
basis~\cite{EGMPR79}, in contrast to covariant gauges where such a
rule is broken. Following \cite{CH94}, we will obtain TMD splitting
functions which complete the set of already available evolution
kernels. Unlike the case of the gluon-to-quark splitting treated in
\cite{CH94}, the resulting splitting kernels have no direct definition
as the coefficient of the BFKL Green's function (or it is equivalent in
the case of $t$-channel quark exchange).  While the TMD quark-to-quark
splitting can be identified as a certain next-to-leading order
contributions to the high-energy resummed non-singlet $P_{qq}$ DGLAP
splitting function, the TMD quark-to-gluon splitting is suppressed by
a power of $x$ w.r.t. the leading logarithmic small-$x$ resummed
$P_{gq}$ DGLAP splitting function.  Nevertheless it is possible to
attempt a definition of such quantities as matrix elements of
reggeized quarks and conventional QCD degrees of freedom in light-cone
gauge.
\\

\begin{figure}[h]
  \centering
  \begin{subfigure}{0.3\textwidth}
    \centering
    \includegraphics{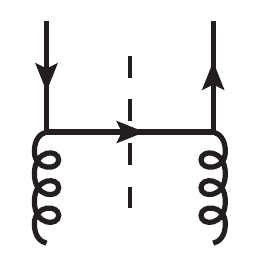}
    \caption{$P_{qg}$}
    \label{fig:Pqg}
  \end{subfigure}%
  ~
  \begin{subfigure}{0.3\textwidth}
    \centering
    \includegraphics{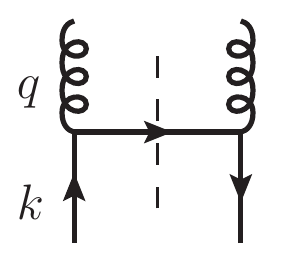}
    \caption{$P_{gq}$}
    \label{fig:Pgq}
  \end{subfigure}%
  ~
  \begin{subfigure}{0.3\textwidth}
    \centering
    \includegraphics{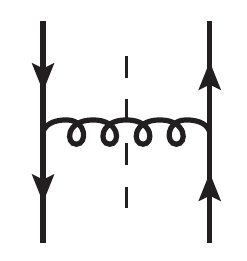}
    \caption{$P_{qq}$}
    \label{fig:Pqq}
  \end{subfigure}%
  \caption{\it Matrix elements for the determination of splitting functions. Lower (incoming) lines carry always momentum $k$, upper (outgoing) lines carry momentum $q$. }
  \label{fig:kinematics}
\end{figure}

Following the framework set by \cite{CFP80,CH94}, the starting point
for the definition of TMD splitting functions requires determination
of the corresponding TMD splitting kernels, 
\begin{align}
  \label{eq:5}
  \hat K_{ij} \left(z, \frac{\kt^2}{\mu^2}, \epsilon, \alpha_s \right) &=
\int \frac{d q^2 d^{2 + 2 \epsilon} {\bm q}}{2 (2 \pi)^{4 + 2
           \epsilon}} \Theta(\mu_F^2 - q^2)  \Pbb_{j,\,\text{in}}
           \otimes \hat{K}_{ij}^{(0)}(q, k) \otimes \Pbb_{i,\,\text{out}}\, .
\end{align}
Here $\hat{K}_{ij}^{(0)}$, $i,j=q,g$ denotes the actual matrix element,
describing the transition of parton $j$ to parton $i$, see
Fig.~\ref{fig:kinematics}, which is defined to include the propagators
of outgoing lines.  In case of gluons, these propagators are taken in
$n\cdot A=0$ light-cone gauge; a similar statement applies to the
polarization of real emitted gluons.  $\Pbb_{i,\,\text{in/out}}$ are
on the other hand semi-projectors on incoming and outgoing lines.
The symbol $\otimes$ represents contraction of indices and
summation. $\mu_F$ denotes the factorization and dimensional
regularization in $d=4+2\epsilon$ dimensions is employed with $\mu^2$
the dimensional regularization scale.
The   Sudakov parametrization for  incoming and outgoing momenta, $k$ and $q$ (see fig.~\ref{fig:kinematics}), reads
\begin{align} \label{eq:kinematics}
  k^\mu & = y p^\mu + k_\perp^\mu,
  &
  q^\mu & = x p^\mu + q_\perp^\mu + \frac{q^2+\qt^2}{2x\sp{p}{n}} n^\mu,
  &
  \qtt & = \qt - z \kt,
\end{align}
with $z=x/y$. 
 The semi-projectors on outgoing lines, $\Pbb_{j,\,\text{out}}$, are directly taken from \cite{CFP80}:
\begin{align}
  \label{eq:2}
   \mathbb{P}_{g,\,\text{out}}^{\,\mu\nu} & = - g^{\mu\nu} 
&
 \mathbb{P}_{q,\,\text{out}} & = \frac{\slashed{n}}{2 \sp{q}{n}}
\end{align}
While outgoing lines are at first treated in 1-1 correspondence to  \cite{CFP80}, the on-shell restriction on incoming lines is now relaxed. The corresponding semi-projectors therefore  require a slight modification. With 
the original projectors  $\Pbb_{j,\,\text{in}}$ of \cite{CFP80},
\begin{align}
  \label{eq:3}
  \mathbb{P}_{g,\,\text{in}}^{\text{\cite{CFP80}}\,\mu\nu} & = \frac{1}{m-2}\left(- g^{\mu\nu} + \frac{k^\mu n^\nu + n^\mu k^\nu}{\sp{k}{n}} \right)\, ,
&
\mathbb{P}^{\text{\cite{CFP80}}}_{q,\,\text{in}} & = \frac{ \slashed{k}}{2}\, ,
\end{align}
are modified to
\begin{align}
  \label{eq:4}
   \mathbb{P}_{g,\,\text{in}}^{\,\mu\nu} & = \frac{k_\perp^\mu k_\perp^\nu}{\kt^2}\, ,
&
 \mathbb{P}_{q,\,\text{in}} & = \frac{y \, \slashed{p}}{2} \,.
\end{align}
While the modified gluon projector has been known since long time
\cite{CH94}, we emphasize that the modified quark projector follows
directly from the decomposition of the high energy projector in
Eq.~\eqref{eq:1}. Its normalization is on the other hand fixed by
requiring agreement with the corresponding projector of \cite{CFP80}
in the collinear limit.  To ensure gauge invariance of the splitting
functions in presence of off-shell momenta, it is further necessary to
modify standard QCD vertices.  The formalism which guarantees that
gauge invariance holds is based on the reggeized quark formalism
\cite{Lipatov:1995pn,Lipatov:2000se,Bogdan:2006af} (for more recent
re-derivation in spin helicity formalism see \cite{HKS13}). The
modification is achieved through adding certain eikonal terms which
then in turn arrange gauge invariance of the vertex. Apart from the
conventional QCD quark-quark-gluon vertex,
$ \Gamma^{\mu}_{qqg}= ig t^a \gamma^\mu$ we have for the off-shell
vertex with one reggeized quark $q^*$
\begin{align}  
   \label{eq:vgq_gen}
  \Gamma_{q^*qg}^\mu (p_{q^*}, p_q, p_g) & = ig t^a \left(\gamma^\mu + \frac{p^\mu}{\sp{p}{p_g}} \slashed{p}_{q^*}\right) &&\text{with} &  p_{q^*} \cdot p &= 0.
\end{align}
Contracting the Lorentz index of this vertex with the gluon momentum
yields $p_{g,\mu} \cdot \Gamma_{q^*qg}^\mu = -ig t^a \slashed{p}_q$
which is equivalent to the corresponding expression for the
conventional quark-quark-gluon vertex if the quark $p_{q}^*$ is taken
on the mass shell. Moreover, in case the second quark is on the
mass shell, we have immediately 
$p_{g,\mu} \cdot \Gamma_{q^*qg}^\mu \bar{u}(p_{q}') = -ig t^a
\slashed{p}_q \bar{u}(p_{q}') = 0$
with $p_{q'}^2 = 0$.
We therefore find that using the generalized vertex Eq.~\eqref{eq:vgq_gen}, the current conservation holds despite of the quark with momentum $p_{q^*}$ being off-shell.

To determine both angular and transverse momentum dependent splitting functions,we start with Eq.~\eqref{eq:5},   perform
color, Dirac and Lorentz algebra, integrate over $q^2$ and shift the
transverse momenta $\qt \to \qtt = \qt - z \kt$, following closely the
treatment in the seminal work of \cite{CH94}. 
We then obtain a set of angular- and transverse momentum dependent splitting functions $\tilde{P}_{ij}$
defined through
\begin{align}
  \label{eq:6angdep}
   \hat K_{ij} \left(z, \frac{\kt^2}{\mu_F^2}, \epsilon, \alpha_s  \right)
     & =
       \frac{\alpha_s}{2 \pi} \, z \,
       \int \frac{\D^{2 + 2 \epsilon}\qtt^2}{\pi^{1 + \epsilon}\mu^{2\epsilon}}  \frac{e^{-\epsilon \gamma_E}}{\qtt^2 }
\notag \\
& \hspace{2cm}
       \Theta\left(\mu_F^2-\frac{\qtt^2 + z(1-z) \kt^2}{1-z} \right)
         \tilde{P}_{ij}^{(0)} \left(z, \kt, \qtt, \epsilon \right)
\end{align}
with the $\overline{\text{MS}}$ scheme coupling $\alpha_s  = \frac{g^2 \mu^{2\epsilon} e^{\epsilon \gamma_E}}{(4 \pi)^{1+ \epsilon}}$
and $\gamma_E$  the Euler-Mascheroni constant. The angular and transverse momentum dependent splitting functions read
\begin{align}
\label{eq:12}
  & \tilde{P}_{gq}^{(0)}\left(z,\kt,\qtt , \epsilon\right)
  = 
  \Cf \left( \frac{\qtt^2}{(\qtt^2 + z(1-z)\kt^2)}\right)^2 \left(\frac{\qtt^2}{ (\qtt - (1-z) \kt )^2}\right)
  \nonumber \\ & \quad
  \times
  \Bigg\{
    \frac{2-2z+z^2}{z} 
    + z(1-z)^2 (1+z^2) \left(\frac{\kt^2}{\qtt^2}\right)^2
    + 4(1 - z)^2\left[
    \frac{\sp{\kt}{\qtt}}{ \qtt^2} 
+  z \left(\!\frac{\sp{\kt}{\qtt}}{\qtt^2}\!\right)^2\right]
  \nonumber \\ & \quad
    + 4z^2(1-z)^2\frac{\sp{\kt}{\qtt} \kt^2}{\qtt^4}
    + 2(1-z)(1+z-z^2) \frac{\kt^2}{ \qtt^2}   
  \Bigg\}
  + \Cf \frac{\epsilon \, z \qtt^2 (\qtt - (1-z)\kt)^2}{(\qtt^2 + z(1-z)\kt^2)^2},
\\
\label{eq:13} 
 & \tilde{P}_{qg}^{(0)}\left(z,\kt,\qtt, \epsilon \right)
  = 
  \Tr \left(\frac{\qtt^2}{ \qtt^2  + z(1-z)\kt^2 }\right)^2
  \nonumber \\ & \qquad \quad
  \times
  \Bigg[1+ 
    4z^2(1-z)^2 \frac{\kt^2}{\qtt^2}
 + 4z(1-z)(1-2z) \frac{\sp{\kt}{\qtt}}{\qtt^2}    
 - 4z(1-z) \left(\!\frac{\sp{\kt}{\qtt}}{\sqrt{\qtt^2 \kt^2}}\!\right)^2
  \Bigg],
\end{align}
\begin{align}
\label{eq:14}
  & \tilde{P}_{qq}^{(0)}\left(z,\kt,\qtt, \epsilon \right)
  = 
  \Cf \left(\frac{\qtt^2}{(\qtt^2 + z(1-z)\kt^2)}\right)^2 \left(\frac{\qtt^2}{(\qtt - (1-z)\kt)^2} \right)
  \nonumber \\ & \quad
  \times
  \Bigg\{
    \frac{1+z^2}{1-z}
    + z^2(1-z) (5-4z+z^2) \left(\frac{\kt^2}{\qtt^2}\right)^2
    - 4z(1-z)^2 \left(\!\frac{\sp{\kt}{\qtt}}{\qtt^2}\!\right)^2
  \nonumber \\ & \quad 
    + 2z(1-2z) \frac{\sp{\kt}{\qtt}}{ \qtt^2}  + 2z(2-7z+7z^2-2z^3) \frac{\kt^2 \sp{\kt}{\qtt}}{\qtt^4}
    + (1+z+4z^2-2z^3) \frac{\kt^2}{ \qtt^2}
\notag 
\\ & \qquad   + \epsilon  (1-z)  \cdot \bigg[1 -2(1-2z) \frac{\kt\cdot\qtt}{\qtt^2}
- 4 z(1-z) \left( \frac{\kt\cdot\qtt}{\qtt^2} \right)^2 + (1 - 2z + 2 z^2) \frac{\kt^2}{\qtt^2}
\notag \\
& \qquad \hspace{4cm}
+z(1-z)(1-2z)
\frac{\kt\cdot\qtt \kt^2}{\qtt^4} + z^2(1-z)^2\left(\frac{\kt^2}{\qtt^2} \right)^2
 \bigg]
  \Bigg\}.
\end{align}
Determination of  both angular and transverse momentum dependent  splitting functions for the splittings quark-to-gluon and quark-to-quark present,  together with the results presented further down in Sec.~\ref{sec:analysis}, the central results of this work.

\section{Gauge invariance of TMD splitting functions}\label{sec:unint-splitt-funct}
The obtained TMD splitting functions will be essential for the
definition of set of TMD evolution equations of TMD parton
distributions. While the above derivation is based on the
2PI-expansion of \cite{CFP80} the derivation might be at first
regarded as not completely satisfactory. While care has been taken to
ensure gauge invariance of the off-shell vertex
Eq.~\eqref{eq:vgq_gen}, the employed formalism heavily relies on the
use of the light-cone gauge and gauge invariance of our result is not
immediately apparent. This is of particular concern, once we relax the
integration over $\tilde{\bm q}$ in Eq.~\eqref{eq:6} to allow for TMD
factorization in the outgoing momentum $\tilde{\bm q}$ and therefore
leave strictly speaking the framework provided by \cite{CFP80}. To
ensure gauge invariance also in this more general case, we will
provide in the following an explicit gauge invariant extension of the
sub-amplitudes Fig.~\ref{fig:kinematics} as well as the projectors. As
a consequence we will both obtain explicitly gauge invariant
sub-amplitudes and verify that any possible gauge dependence hidden in
the propagators of the outgoing parton with momentum $q$ and/or the
real produced parton with momentum $p'=k-q$ will cancel. In
particular, while calculations are no longer restricted to the
light-cone gauge as in Sec.~\ref{sec:splitt-kern-from}, they agree at
every stage precisely with the results derived in this gauge.  To this
end we first generalize the projector of the outgoing gluon in
Eq.~\eqref{eq:2}. Another source of potential violation of gauge
invariance is due to the use of explicit cut-offs in
Eq.~\eqref{eq:5}. A generalization of our results to a
cut-off-independent formulation is left at this stage as a task for
future research, restricting ourselves for the time being to the
proper definition of gauge-invariant sub-amplitudes.
\\

With the polarization tensor of the gluon propagator
in the light-cone gauge given by
\begin{equation}
  \label{eq:2_projector}
 \Delta_{\mu\mu'} (q) = - g_{\mu\mu'} + \frac{q^{\mu'} n^\mu + n^{\mu'}q^\mu}{q\cdot n},
\end{equation}
we define the new projector
\begin{align}
  \label{eq:3_new}
\tilde{\mathbb{P}}_{g,\,\text{out}}^{\,\mu\nu}(q,n) & \equiv 
  \Delta_{\mu\mu'} (q) \mathbb{P}_{g,\,\text{out}}^{\,\mu'\nu'} \Delta_{\nu'\nu}(q) = -g_{\mu\nu} + \frac{q^\mu n^\nu + n^\mu q^\nu}{q\cdot n} - q^2 \frac{n^\mu n^\nu}{(q \cdot n)^2} ,
\end{align}
which fulfills the following properties:
\begin{align}
  \label{eq:7}
   0 &=  \tilde{\mathbb{P}}_{g,\,\text{out}}^{\,\mu\nu} \cdot q_\mu 
= 
  \tilde{\mathbb{P}}_{g,\,\text{out}}^{\,\mu\nu} \cdot q_\nu 
=
  \tilde{\mathbb{P}}_{g,\,\text{out}}^{\,\mu\nu} \cdot n_\mu
=
  \tilde{\mathbb{P}}_{g,\,\text{out}}^{\,\mu\nu} \cdot n_\nu \, .
\end{align}
Furthermore
\begin{align}
  \label{eq:8}
    \tilde{\mathbb{P}}_{g,\,\text{out}}^{\,\mu\nu} \cdot q_\perp^\mu q_{\perp}^\nu = {\bm q}^2
\end{align}
and hence the combination
\begin{align}
  \label{eq:9}
    \tilde{\mathbb{P}}_{g,\,\text{out}}^{\,\mu\nu}(q,n)  \tilde{\mathbb{P}}_{g,\,\text{in}}^{\,\mu'\nu'} (q_\perp), 
\end{align}
is indeed a projector. Due to the properties Eq.~\eqref{eq:7}, one also has
\begin{align}
  \label{eq:10}
   \tilde{\mathbb{P}}_{g,\,\text{out}}^{\,\mu\nu'}(q,n) \Delta_{\nu'\nu}(q) =  \tilde{\mathbb{P}}_{g,\,\text{out}}^{\,\mu\nu'}(q,n) =  \Delta_{\mu\mu'}(q) \tilde{\mathbb{P}}_{g,\,\text{out}}^{\,\mu'\nu}(q,n)\,.
\end{align}
Using therefore Eq.~\eqref{eq:3_new} in the analysis of the previous
section, will leave our results unchanged. The second modification
concerns the sub-amplitudes Fig.~\ref{fig:kinematics}. In the high energy  limit, corresponding gauge invariant vertices can be easily
derived within the reggeized quark formalism. To ensure gauge invariance in presence of both off-shell momenta $k$ and $q$, with $q$
of the general form Eq.~\eqref{eq:kinematics}, these vertices require
a slight generalization, similar to the one employed already in
\cite{Hautmann:2012sh}. The version to be used in the following reads
\begin{align}  
  \label{eq:vqg}
  \Gamma_{q^*g^*q}^\mu (q, k , p') & = igt^a \left(\gamma^\mu - \frac{n^\mu}{\sp{k}{n}} \slashed{q}\right),
  \\ 
  \label{eq:vgq}
  \Gamma_{g^*q^*q}^\mu (q, k, p') & = igt^a \left(\gamma^\mu - \frac{p^\mu}{\sp{p}{q}} \slashed{k}\right),
  \\
  \label{eq:vqq}
  \Gamma_{q^*q^*g}^\mu (q, k, p') & =igt^a \left( \gamma^\mu - \frac{p^\mu}{\sp{p}{p'}} \slashed{k} + \frac{n^\mu}{\sp{n}{p'}} \slashed{q}\right).
\end{align}
where we used $p' = k-q$ for the momentum of the real produced particle and $q^*$, $g^*$indicate an off-shell quark and gluon; the momentum  $k$  and momentum $q$ refer always to incoming and outgoing particles respectively. In particular, these vertices obey
\begin{align}
  \label{eq:11}
 {k}_\mu \cdot  \Gamma^{\mu}_{q^*g^*q}(k,q) \bar{u}(p') & = 0 
&
{q}_\mu \cdot  \Gamma^{\mu}_{q^*q^*g}(k,q) \bar{u}(p') & = 0 
&
{p'}_\mu \cdot  \Gamma^{\mu}_{q^*q^*g}(k,q) & = 0 \, .
\end{align}
Due to these properties, any gauge dependence induced by either the
polarization tensor of a $t$-channel gluon with momentum $q$, a real
produced gluon with momentum $p' = k-q$, or a $t$-channel gluon with
momentum $k$ is canceled and the overall result is gauge-invariant. In
particular it is trivial to check that the results obtained in the
previous section using the light-cone gauge, generalize directly to
the present formulation. A last comment is in order concerning the
universality of our results. As pointed out in the beginning of
Sec.~\ref{sec:splitt-kern-from}, unlike the splitting function of
\cite{CH94}, our splitting functions cannot be uniquely associated
with the e.g. next-to-leading order coefficient of the small-$x$
gluon Green's function etc. Indeed there will be always contributions
of similar order of magnitude than elements of our splitting
functions, which are not contained in its definition. Our splitting
functions comprise however a set of contributions which
\begin{itemize}
\item reduces in the collinear limit
to collinear splitting functions
\item reduces in the high energy  limit
to corresponding high energy factorized  expressions (guaranteed through the use of
the reggeized quark and gluon vertices) 
\item combines both limits in a gauge invariant way.
\end{itemize}
It is then the combination of these three requirements which provides
strong constraints on the terms contained in the definition of our
TMD splitting functions.

\section{Angular averaged TMD splitting functions and singularity structure}
\label{sec:analysis}

In the following section we further analyze our results of Sec.~\ref{sec:splitt-kern-from}.
While the explicit angular-momentum-dependence of our results might be of interest for further Monte-Carlo realizations which aim at description of exclusive final states, the evolution of TMD parton distribution functions generally requires only angular-averaged splitting functions. Furthermore, the splitting functions turn out to be divergent in certain regions of phase space, which will be identified below. 

\subsection{Angular averaged TMD splitting functions}
\label{sec:averagedTMD}

To arrive at a result similar to the one obtained in  \cite{CH94} for the TMD $P_{qg}$, it is further necessary to average over the azimuthal angle.
With
\begin{align}
  \label{eq:6}
   \hat K_{ij} \left(z, \frac{\kt^2}{\mu_F^2}, \alpha_s , \epsilon \right)
     & =
     \frac{\alpha_s}{2 \pi} \, z \!\!\! \int\limits_0^{(1-z)(\mu_F^2-z\kt^2)} \frac{\D\qtt^2}{\qtt^2}
     \left( \frac{\qtt^2}{\mu^2} \right)^\eps  \frac{e^{-\epsilon \gamma_E}}{\Gamma(1 + \epsilon)}
     {P}_{ij}^{(0)} \left(z, \frac{\kt^2}{\qtt^2} , \epsilon\right),
\end{align}
which then defines the TMD splitting functions $ P_{ij}$, we reproduce for the gluon-to-quark splitting the result of \cite{CH94}, also calculated in \cite{Ciafaloni:2005cg,Hautmann:2012sh}
\begin{align} \label{eq:pqg}
  P_{qg}^{(0)} \left(z, \frac{\kt^2}{\qtt^2}, \epsilon \right)
  & = 
  \Tr \left(\frac{\qtt^2}{\qtt^2 + z(1-z)\,\kt^2}\right)^2
  \Bigg[
    z^2+(1-z)^2
    +
    4 z^2 (1-z)^2 \frac{\kt^2}{\qtt^2}
  \Bigg]. 
\end{align}
For the  new TMD splitting functions  we obtain 
\begin{align} \label{eq:pgq}
  {P}_{gq}^{(0)} \left(z, \frac{\kt^2}{\qtt^2}, \epsilon \right)
  & =
 C_F  \Bigg[ \frac{2 \qtt^2 }{z |\qtt^2 - (1-z)^2 \kt^2|}
  \nonumber \\ &
\hspace{2cm} -
  \frac{\qtt^2 (\qtt^2(2-z) + \kt^2 z(1-z^2)) - \epsilon z (\qtt^2 + (1-z)^2 \kt^2)}{(\qtt^2 + z(1-z)\kt^2)^2}
  \Bigg]
  ,
\\ \label{eq:pqq}
  {P}_{qq}^{(0)} \left(z, \frac{\kt^2}{\qtt^2}, \epsilon \right)
  & = 
\Cf \left(\frac{\qtt^2}{ \qtt^2 +
    z(1-z)\kt^2} \right) \bigg[ \frac{\qtt^2 + (1-z^2)\kt^2}{(1-z)|\qtt^2 - (1-z)^2 \kt^2|} 
\notag \\
& \hspace{1.3cm}
+
\frac{z^2 \qtt^2 -z(1-z)(1 - 3z + z^2)\kt^2 + (1-z)^2\epsilon (\qtt^2 + z^2 \kt^2)}
                              {(1-z)  (\qtt^2 +    z(1-z)\kt^2)}\bigg]\, .
\end{align}
As expected from our method to construct TMD splitting functions, we obtain in the collinear limit ($\kt^2/\qtt^2 \to
0$) the  well-known real parts of the leading-order Altarelli-Parisi splitting functions in $d = 4 + 2\epsilon$ dimensions:
\begin{align}
  P_{gq}^{(0)}(z,0, \epsilon)
  & = 
\Cf \frac{1+(1-z)^2 + \epsilon z^2}{z},
\\
  P_{qg}^{(0)}(z,0, \epsilon)
  & = 
\Tr \left(z^2 + (1-z)^2\right),
\\
  P_{qq}^{(0)}(z,0, \epsilon)
  & = 
\Cf\, \frac{1+z^2 + \epsilon (1-z)^2}{1-z}.
\end{align}

\subsection{Singularity structure of the TMD splitting functions}
\label{sec:sing_struct}

Unlike the $P_{qg}$ TMD splitting function, the splitting functions in
Eq.~\eqref{eq:pqg} and Eq.~\eqref{eq:pqq} develop singularities in
certain regions of phase space. These singularities can be organized
into two groups: those associated with the limit $z \to 1$, only
present for the splitting $P_{qq}$, and those associated with the
limit $|\qtt| \to (1-z)|\kt|$, present for both $P_{gq}$ and
$P_{qq}$.  The coefficient of the $z\to 1$ singularity reads
\begin{align}
  \label{eq:15}
\lim_{z \to 1} P_{qq}^{(0)}\left(z, \frac{\kt^2}{\qtt^2} \right) &= \Cf \frac{2}{1-z} 
\end{align}
and coincides with the  $z\to 1$ singularity  of
the conventional collinear splitting functions where it is known to be
regularized by corresponding virtual corrections to the splitting
kernel. We expect a similar mechanism to be realized in the case of
the $P_{qq}$ splitting kernel with full transverse momentum
dependence. The nature of the second singularity is more intriguing,
since it is present for both diagonal ($P_{qq}$ ) and off-diagonal
($P_{gq}$) splitting kernels. The coefficients of this singularity is provided by
\begin{align}
  \label{eq:16}
\lim_{|\qtt| \to (1-z) |\kt|} P_{gq}^{(0)}\left(z, \frac{\kt^2}{\qtt^2}, \epsilon \right) 
 & =
\frac{2 \Cf}{z} \frac{ \qtt^2}{|\qtt^2 - (1-z)^2 \kt^2|} 
 \notag \\
\lim_{|\qtt| \to (1-z) |\kt|} P_{qq}^{(0)}\left(z, \frac{\kt^2}{\qtt^2}, \epsilon \right) 
 & = 
\frac{2 \Cf}{1-z} \frac{ \qtt^2}{|\qtt^2 - (1-z)^2 \kt^2|} 
\end{align}
For the $P_{qq}$ splitting function, this singularity always overlaps with the $z\to 1$ singularity. At the level of the angular-dependent splitting kernels
Eq.~\eqref{eq:12} and Eq.~\eqref{eq:14}, this singularity is easily
identified with the vanishing of the transverse momentum of the real,
emitted parton i.e.  of the real gluon ($P_{qq}$) and of the
real quark ($P_{gq}$) respectively. To analyze the precise structure of the singularities within dimensional regularization it is convenient to switch to the re-scaled momentum $\pt = \frac{\kt - \qt}{1-z}$ instead of $\qtt$. We then obtain
\begin{align}
  \label{eq:17qq}
\hat K_{qq} \left(z, \frac{\kt^2}{\mu_F^2}, \alpha_s \right)&=  \frac{\alpha_s}{2 \pi} z \int \frac{d^{2 + 2\epsilon} \pt}{\pi^{1+\epsilon}\mu^{2\epsilon}} 
\, e^{-\epsilon \gamma_E} \, 
\Theta\left(\mu_F^2 - (1-z) (\pt - \kt)^2 - z \kt^2 \right)
\notag \\
& \hspace{1.6cm}\Bigg\{
 \frac{1}{(1-z)^{1 - 2 \epsilon}} \left(\frac{1}{\pt^2} + \frac{\kt^2 + z(1-z)\pt^2}{\pt^2 [(1-z) (\pt - \kt)^2 + z \kt^2]} \right) 
\notag \\
&  \hspace{1.6cm}-
\frac{\pt^2 z(1-z)^{2 + 2\epsilon} (1 + \epsilon)}{ [(1-z) (\pt - \kt)^2 + z \kt^2]^2}
+ 
\frac{\epsilon (1-z)^{1 + 2\epsilon}}{ (1-z) (\pt - \kt)^2 + z \kt^2}
\Bigg\}
\end{align}
\begin{align}
  \label{eq:17gq}
\hat K_{gq}\left(z, \frac{\kt^2}{\mu_F^2}, \alpha_s \right)&=  \frac{\alpha_s}{2 \pi} z \int \frac{d^{2 + 2\epsilon} \pt}{\pi^{1+\epsilon} \mu^{2 \epsilon}}\, e^{-\epsilon \gamma_E} \,  \Theta\left(\mu_F^2 - (1-z) (\pt - \kt)^2 - z \kt^2 \right)
\notag \\
& \hspace{1.6cm}\Bigg\{
 \frac{(1-z)^{2\epsilon}}{z} \frac{2}{\pt^2} 
+ \frac{\pt^2 z(1-z)^{2 + 2\epsilon} (1 + \epsilon)}{ [(1-z) (\pt - \kt)^2 + z \kt^2]^2} \notag \\
&  \hspace{6.2cm}
-
\frac{2 (1-z)^{1 + 2\epsilon}}{ (1-z) (\pt - \kt)^2 + z \kt^2}
\Bigg\}
\end{align}
It is now possible to isolate the singularities of interest using a phase space slicing parameter $\lambda \to 0$ which splits the 
integration over $\pt$ into regions  $|\pt| < \lambda$, $|\pt| > \lambda$.
Defining\footnote{Note that the kernels $K_{qq}^{(0)\text{fin.}}$ and  $K_{gq}^{(0)\text{fin.}}$ still contain divergences due to the $z \to 1$ singularity} $K_{qq}^{(0)\text{fin.}}$ and  $K_{gq}^{(0)\text{fin.}}$ as the kernels given in Eq.~\eqref{eq:17qq} and Eq.~\eqref{eq:17gq}, but with the integration measure $d^{2+2 \epsilon} \pt$ replaced appropriately by $d^{2+2 \epsilon} \pt \cdot \Theta(\pt^2 - \lambda^2)$ and $d^{2+2 \epsilon} \pt \cdot \Theta(\lambda^2-\pt^2)$ we have 
\begin{align}
  \hat K_{qq} \left(z, \frac{\kt^2}{\mu_F^2}, \alpha_s , \epsilon\right)
&=  
 \hat K_{qq}^{\text{fin.}} \left(z, \frac{\kt^2}{\mu_F^2}, \alpha_s , \epsilon\right)
 + 
 \hat K_{qq}^{\text{div.}} \left(z, \frac{\kt^2}{\mu_F^2}, \alpha_s, \epsilon \right) 
\notag \\
  \hat K_{gq} \left(z, \frac{\kt^2}{\mu_F^2}, \alpha_s, \epsilon \right)
&=   
\hat K_{gq}^{\text{fin.}} \left(z, \frac{\kt^2}{\mu_F^2}, \alpha_s, \epsilon \right) +
\hat K_{gq}^{\text{div.}} \left(z, \frac{\kt^2}{\mu_F^2}, \alpha_s, \epsilon \right)
\end{align}
with
\begin{align}
  \label{eq:17div}
\hat K_{qq}^{\text{div.}} \left(z, \frac{\kt^2}{\mu_F^2}, \alpha_s, \epsilon \right) 
&=
\frac{\alpha_s}{ \pi} \Theta(\mu_F^2 - \kt^2) \frac{e^{-\epsilon \gamma_E}}{\Gamma(1 + \epsilon)} \frac{\lambda^{2\epsilon}}{\epsilon (1-z)^{1- 2 \epsilon}} + \mathcal{O}(\lambda) \notag \\
& \hspace{-2cm} =  \frac{\alpha_s}{ \pi} \Theta(\mu_F^2 - \kt^2)  \frac{e^{-\epsilon \gamma_E}}{\Gamma(1 + \epsilon)} 
 \left(\frac{\lambda^2}{\mu^2} \right)^{\epsilon} 
\frac{1}{\epsilon} \left(\frac{1}{2 \epsilon} \delta(1-z) + \frac{1}{(1-z)_+^{1 - 2\epsilon}} \right)+  \mathcal{O}(\lambda) 
\notag \\
\hat K_{gq}^{\text{div.}} \left(z, \frac{\kt^2}{\mu_F^2}, \alpha_s \right) 
&=
\frac{\alpha_s}{ \pi} \Theta(\mu_F^2 - \kt^2) \frac{e^{-\epsilon \gamma_E}}{\Gamma(1 + \epsilon)}
 \left(\frac{\lambda^2}{\mu^2} \right)^{\epsilon}
 \frac{\left((1-z)\right)^{2\epsilon}}{\epsilon z} + \mathcal{O}(\lambda)
\end{align}
where we made use of the limit $\lambda \to 0$. We furthermore introduced the  usual plus-prescription
\begin{align}
  \int_0^1  d z\frac{f(z)}{(1-z)_+} & \equiv  \int_0^1  d z\frac{f(z)- f(1)}{(1-z)},
\end{align}
and made use of the identity
\begin{align}
  \frac{1}{(1-z)^{1-2\epsilon}} & = \frac{1}{2 \epsilon} \delta(1-z) + \frac{1}{(1-z)_+^{1-2\epsilon}} 
\notag \\
& =  \frac{1}{2 \epsilon} \delta(1-z) + \frac{1}{(1-z)_+} + 2 \epsilon \left( \frac{\ln (1-z)}{(1-z)}\right)_+ + \mathcal{O}(\epsilon^2)\, .
\end{align}
Note that since the real emitted particle is on-shell, the
vanishing of its transverse momentum $\pt$ implies also vanishing of
the component parallel to $n$. As a consequence the momentum of the
emitted particle is in this case collinear to the initial proton
momentum $p$. For a hands-on approach, it appears therefore to be
natural to avoid this singularity by introducing a cut-off, similar to
$\lambda$ in the $K_{qq}^{(0)\text{fin.}}$ and
$K_{gq}^{(0)\text{fin.}}$ terms or by imposing an angular ordering
inspired constrained on the $t$-channel momenta such as
$|\qtt|/(1-z) > |\kt|$ which avoids the singular region.  Such a
treatment would then allow for first numerical tests of the proposed
TMD splitting functions and for their application to phenomenological
studies.  A complete theoretical treatment of this singularity would on
the other hand require the determination of virtual corrections (in
the case of the $P_{qq}$ splitting) and most likely the realization of
a systematic subtraction mechanism which removes parton emission
collinear to the initial proton momentum from the TMD splitting
kernels. Both tasks are beyond the scope of this work and are left as
a task for future research.

\section{Summary and Outlook}
\label{sec:conclusions-outlook}
In this paper we extended the method developed by Catani and Hautmann
for the determination of transverse-momentum-dependent parton
splitting functions to splittings of initial $k_T$-dependent quarks,
based on factorization of cross-sections in the high energy
limit. Gauge invariance of underlying amplitudes in presence of
off-shell partons is achieved due to the reggeized quark calculus,
which supplements conventional QCD vertices by certain eikonal
contributions. While our approach is heavily based on the 2PI
expansion in the light-cone gauge by Curci et al., we have been
able to verify that it is possible to generalize the employed
projectors in a way,  such that the choice of gauge for the
sub-amplitudes, which underlie the derivation of our splitting kernels,
becomes irrelevant i.e. our TMD splitting kernels are
independent of the employed gauge. While our splitting kernels are in
this way well defined objects, there are not necessarily universal,
since they cannot be directly defined as the coefficients of
e.g. the high energy resummation of a certain TMD parton
distribution function, such as the TMD gluon-to-quark splitting
functions. They are merely constrained by the requirement to reduce in
the collinear and
high energy  limit to the well-known exact expressions.\\

The current study determines only the real contribution to the TMD
quark-to-quark and quark-to-gluon splitting kernels. Future studies
will have to focus on the determination of the corresponding virtual
corrections for the TMD quark-to-quark splitting function, the
development of a coherent framework which allows for a systematic
subtraction of singularities not canceled by virtual corrections and
finally the formulation of appropriate coupled evolution equations for
TMD parton distribution functions. As a long term goal, a matching of
TMD evolution based on factorization in the soft-collinear limit, see
e.g. \cite{Collins:2014jpa,Echevarria:2012pw, Becher:2010tm} is
a task which needs to be addressed.

\section*{Acknowledgments}
Useful discussions with  Dimitri Colferai and  Andreas van Hameren are acknowledged.
O.G. and K.K. would like to acknowledge support by the Polish National
Science Center with Grant No.
DEC-2013/10/E/ST2/00656. M.H. acknowledges support by
UNAM-DGAPA-PAPIIT grant number 101515 and CONACyT-Mexico grant number
128534 and CB-2014-22117.

\bibliography{main}

\end{document}